\documentstyle[12pt]{article}

\def\onehalfspace{\baselineskip=18pt}

\begin{document}

\onehalfspace

\begin{center}
{\large {\bf Neural Networks and the Classification of\\Active Galactic 
Nucleus Spectra}}
\end{center}

\begin{center}
 Daya M. Rawson$^1$, Jeremy Bailey$^2$ \& Paul J. Francis$^3$  
\\$^{1}$Mount Stromlo and Siding Springs Observatory, Australian National 
University, Private Bag, Weston Creek, ACT, 2611, Australia. \\ 
E-mail: daya@merlin.anu.edu.au
\\$^{2}$Anglo-Australian Observatory, PO Box 296, Epping, NSW 2121, 
Australia. \\ E-mail:
jab@aaoepp.aao.gov.au
\\$^{3}$School of Physics, University of Melbourne, Parkville, 
ictoria 3052, Australia. E-mail:
pfrancis@physics.unimelb.edu.au 
\end{center}

\medskip

\noindent
{\bf Abstract:}
The use of Artificial Neural Networks (ANNs) as a classifier of digital 
spectra is investigated. Using both simulated and real data, it is shown 
that neural networks can be 
trained to discriminate between the spectra of different classes of active
galactic nucleus (AGN) with realistic sample sizes and signal-to-noise ratios.
By working in the Fourier domain, neural nets can classify objects without
knowledge of their redshifts.

\medskip

\noindent
{\bf keywords:} galaxies: nuclei, galaxies: quasars: general, \\ 
galaxies: seyferts, methods: data analysis

\medskip

\noindent {\bf 1. Introduction}

\medskip

The advent of current generation spectrographs, such as the 2dF system
on the AAT (Taylor, 1995) is leading to order-of-magnitude increases in the
number of spectra that can be obtained in a given amount of observing time. 
This in turn has stimulated research into computer algorithms capable of 
handling large samples of spectra (eg. Francis et al. 1992).

One family of such algorithms are artificial neural networks. These 
algorithms,
inspired by a model of the workings of the human brain, have recently found
application in fields as diverse as sonar signal processing (Gorman \&
Sejnowski 1988), and medical diagnostics (Rayburn et al. 1991). Neural
networks are increasingly being used in astronomy (eg. Storrie-Lombardi
et al., 1992, Lahav et al., 1994, von Hippel et al. 1994, and 14 papers
in a special issue of Vistas in Astronomy, 1994, 38, 251).

In their most common application, artificial neural networks function as
classification algorithms. They have one unique feature that sets them
apart from most classification algorithms: they can be trained. The procedure
is known as supervised learning. To train a neural network to divide some 
large dataset up into different classes of object, one would classify
some subset of the whole sample manually. The neural net is then trained using
this subset to identify the various classes of object in the data. It can
then be used to automatically classify the remainder of the dataset. Unlike
most classification algorithms, this trainability allows neural networks
to duplicate the subjectivity so common in astronomical classification.

The aim of this paper is to determine if neural networks can
be usefully applied to data sets of realistic astronomical spectra, with
typical sample sizes, redshift ranges and signal-to-noise ratios.
This is tested using both real and simulated spectra of active
galactic nuclei (AGN).

\medskip

\noindent {\bf 2. Method}

\medskip

The simplest and most widely used type of neural network algorithm
is backpropagation, and this is the algorithm used in this
paper. Excellent descriptions of neural networks and the backpropagation
algorithm can be found in Hertz, Krogh \& Palmer (1991), Maren, Harston 
\& Pap (1990), Hinton (1992), van Camp (1992), and Lahav et al. (1995).
A neural network is defined by a series of variables known as weights.
The process of training consists of choosing the weights such that
the network duplicates the known classifications of some training set of 
data. Once the weights are chosen, the neural network can then be
applied to other, previously unclassified data sets. The backpropagation
algorithm is one of many iterative techniques for choosing the weights.

In this paper, a modified version of the code of Rao \& Rao (1993)   
was used; the code (in C or C++) is available on request from DMR.
Any backpropagation code using Sigmoid transfer functions will work
equally well. A full technical description of the algorithm used can
be found in Rawson (1994).

Many classification schemes for AGN spectra use derived quantities as
their input variables, such as emission-line fluxes and continuum slopes
(eg. Boroson \& Green 1992). 
Unfortunately, the measurement of these derived quantities from spectra
introduces an inevitable subjectivity into the classification (Francis et
al. 1992). In this paper, a different approach was tried; the spectra are 
used directly, with the flux in each wavelength bin an input variable. This 
is more objective, but greatly increases the computational load.

\medskip

\noindent {\bf 3. Simulations}

\medskip

The neural network was first tested with a synthetic data set, to
investigate its sensitivity to noise, sample size and redshift. The
classification of AGN spectra into Seyfert I and Seyfert II
galaxies (Osterbrock 1989) was chosen as a test of our neural network.
Seyfert I and II galaxies differ primarily in the velocity width of their 
permitted emission lines.

The synthetic spectra covered the rest-frame wavelength region 4800--5100 \AA .
Three emission-lines were included; H-$\beta$ and the [O~III] doublet. No
continuum emission was included. The peak fluxes of the lines were fixed for 
the H-$\beta$ and [OIII] doublet in the ratio $2:3:9$ respectively. The 
[O~III] lines were assigned velocity widths of $800\ {\rm km\ s}^{-1}$ 
(Full Width at Half Maximum Height, FWHM). The velocity
width of H-$\beta$ was randomly assigned a value in the range
$100 < {\rm FWHM} < 4500 {\rm km\ s}^{-1}$, the widths being drawn from
a uniform distribution. Peak flux was conserved with a magnitude of 10 
as the width was changed. Objects with H-$\beta$ FWHM $> 1500 {\rm km\ s}^{-1}$
were considered to be Seyfert I galaxies; all others were Seyfert II galaxies.
The sample was divided equally between the two classes. 
Random noise was added to the spectra; the noise being drawn from a Gaussian
distribution, with a standard deviation independent of wavelength. 
The signal-to-noise ratio in the emission-lines, for a noise standard 
deviation of $1$, is comparable to that of a typical AGN spectrum with a 
continuum signal-to-noise ratio of $\sim 10$ (Francis et al. 1991), ie. 
typical of spectral surveys.

100 simulated spectra were generated for use as a training set; this size
was chosen to match the typical sizes of many astronomical samples. Each 
spectrum was composed of 300 points at 1 \AA\ intervals from 4800\AA\ to 
5100\AA, with each point corresponding to the input to one neuron in the 
first layer.

A two-level neural network was initially used. This is only capable of
linear classification; it defines a plane in the $n$-dimensional decision 
space defined by the input variables (in this case the fluxes in the
300 spectral bins), and classifies objects as Seyfert I or Seyfert II
galaxies depending upon which side of the plane they lie (a 300:1
perceptron network).

The network was trained on this set, using the backpropagation algorithm.
Training continued until the network would reliably match the known
classification. Approximate training times varied between 10 and 20 minutes 
on a Sun sparc10.

Once the neural network had been trained on the training set, it was
applied to an additional set of 1000 simulated spectra (the testing set), 
to measure the success rate of the classification.
Figure~1 shows the fraction of Seyferts classified correctly as a 
function of the noise added to the simulations. The results are
encouraging; for noise levels of $< 2$ (signal-to-noise ratios of
$\sim 5$ or better), the classification is more than 90\% accurate. 
The classification is still 70\% correct for noise standard deviations 
as large as $3.5$ (signal-to-noise ratios $\sim 3$ per \AA ).
{\em This demonstrates that the simple neural network backpropagation 
algorithm is
capable of  classifying spectra with typical noise levels and
sample sizes.}

In these simulations, the noise level in the training set and in the 
testing set were identical.
One possible strategy, however, would be to train the 
network upon data of better quality than that which it is to be used 
to classify. To simulate this, the neural network was trained 
on a noise-free training set (still of 100 objects). Remarkably, this
was found to {\em degrade} the performance of the neural network at 
classifying noisy data. The likely explanation is as follows: in the 
$N$-dimensional space defined by the 300 input variables, a noise-free
training set will be a curved line, along which the width of H-$\beta$
increases. The training process chooses a plane which separates
Seyfert-I and Seyfert-II galaxies, but in the absence of noise, this
is a degenerate problem, as any plane cutting the curved line at the
appropriate place will do equally well. Such a plane may not lie 
perpendicular to the line it bisects. When noise is added, the line becomes
fuzzy, and if the classification plane is not parallel, many objects
may be misclassified. Training the neural network on noisy data forces
the classification plane to lie perpendicular to the line of sample
points.

\newpage

\noindent {\it 3a. Redshift Dependence}

\medskip

In the above simulations, it was tacitly assumed that the redshifts of
the AGNs are known. In practice, we would like the classification algorithm
to work without the need to be told a redshift.

As a first attempt to model this, the simulated spectra were randomly
assigned redshifts in the range $0 < z < 0.1$, and 
the spectra were shifted appropriately. The neural network was once again 
trained on a training set of 100 spectra, and tested on a testing set of
1000 spectra.
Even for noise-free data, the success-rate  was only $\sim 45$---55\%, 
no better than random. This is because
putting the spectra at a range of redshifts, thereby changing both the 
line-width and position, makes the boundary in decision
space between Seyfert I and Seyfert II galaxies non-linear.

The standard solution to this problem is to add an additional
layer (a hidden layer) of neurons to the neural network (Maren, Harston 
and Pap 1990). The extra layer of neurons is equivalent to combining
several two-layer neural networks; instead of a single decision plane in
the decision space defined by the 300 input variables, the classification can
be done on a more complex surface, made up of several planes. This in
principle allows the neural network to handle non-linear classification.

In practice, however, adding a hidden layer to the neural network did not
significantly improve the classification success rate, regardless of the 
size of hidden layer employed. We hypothesise that
this is due to the inadequate size of the training set; $\sim 100$
spectra is not sufficient to define the complex boundary between
Seyfert I and Seyfert II spectra at a range of redshifts in the decision
space.

As an alternative, the classification was attempted in the Fourier domain.
A decision space made up of Fourier amplitudes should be invariant
under changes in redshift. A Fast Fourier Transform routine from Press et 
al.(1992). was used, and the resulting 300 Fourier components were used
as inputs into a two-layer neural network (a 300:1 perceptron). 

This technique performed well; success rates as a function of noise
level are shown in Figure~1. For signal-to-noise ratios $\sim 10$,
success rates are comparable to those obtained when the redshift is
known. The performance does, however, decline for noiser data.

In real data, the continuum emission could complicate things. Unless
removed or smoothed, edge effects will confuse the Fourier analysis.
Standard techniques used in cross-correlation redshift measurement,
such as cosine bell weighting, would introduce features into the
Fourier transform on wavelength scales comparable to the emission-line
widths which define the spectral classes. We therefore suggest that
fitting and subtracting a continuum may be necessary. 

Neural networks are thus capable of successfully classifying data for
which redshifts are not known. For data with low signal-to-noise ratios,
however, redshift information does improve the classification.

\medskip

\noindent {\it 3b. Training Sample Size}

\medskip

Sensitivity of the performance of a neural network to the size the 
training set is particularly important because large samples of 
pre-classified spectra are difficult to obtain. The simulations
described above were therefore run repeatedly, varying the number of
spectra in the training set. Results are shown in Figure~2.

In all cases, a sample of $\sim 100$ spectra seems adequate; increasing 
the training set beyond this size does little to improve the 
success rate of the classification.

Using a larger training set does however give a more {\em reproducible}
classification. This was tested by repeating the classification using a
different randomly generated training set of the same size. For training
set sizes of $\sim 50$, $\sim 5$\% of objects in the testing set would 
change classification when different training sets were used. This
fraction dropped to $\sim 2$\% for training set sizes $\sim 1000$. We
hypothesise that the larger training set allows the decision space boundary 
to be better defined.

Thus for most purposes, a training set of $\sim 100$ spectra is adequate, 
unless high levels of reproducibility are required.
 
\medskip

\noindent{\it 3c. Classifying Real Data}

\medskip

As a final test, the neural network was applied to a sample of real
data; QSO spectra from the Large Bright QSO Survey (LBQS).
The LBQS is a large optically selected QSO sample, containing 1054
objects. References to the LBQS can be found Hewett, Foltz \& Chaffee 
(1995). LBQS
QSOs span the redshift range $0.2 < z < 3.4$; the spectra have resolutions
of $\sim 10$\AA \ and uniform signal-to-noise ratios of $\sim 10$ per
resolution element.

Roughly 10\% of the QSOs in the LBQS belong to the sub-class of QSOs known
as broad absorption-line QSOs (BALQSOs). This subset are characterised by
P-Cygni type profiles to some of their high ionisation emission-lines,
evidence for massive high velocity out-flowing winds. The properties of the
BALQSOs in the LBQS are discussed by Weymann et al. (1991), and Francis,
Hooper \& Impey (1993). Broad absorption-lines are only seen in ultra-violet
resonance lines, particularly C IV, which can only be seen in the spectra of
higher redshift QSOs: we therefore restrict our analysis to QSOs with $z>1.7$.

The aim of this test was to measure the robustness of the neural network
algorithm when faced with real data, by training it to classify QSOs in 
the LBQS as BALQSOs or not on the basis of their spectra. The network
was trained to reproduce the `by eye' classification of a human
expert (Craig Foltz).

Accurate redshifts are known for the LBQS QSOs, so the classification
could be done in rest-frame wavelengths; the Fourier transform technique 
was not required. The spectra were shifted to the rest-frame, and rebinned 
into 16.3 \AA\ bins to minimise the number of input variables. All spectra
were normalised to unit mean flux, so that the analysis was sensitive
only to the shape of the spectra, not their normalisation.

The imbalance in numbers between BALQSOs and non-BALQSOs was found to
be a problem. It had the effect that the weights were trained to recognise 
only the more common non-BALQSOs type. All changes in the weights to 
facilitate the classification of the BALQSOs were overwhelmed by the number 
of times the weights were changed for the classification of non-BALQSOs,
delaying or preventing convergence of the training. This was overcome
by duplicating the BALQSOs in the training set, so that their
numbers matched the non-BALQSOs. BALQSOs and non-BALQSOs were alternated
in the training set.

Using a training set of $\sim 250$ QSOs, 33 of which were BALQSOs, the
network was successfully trained. The trained neural network achieved
a success rate of $\sim 90$\% , comparable to that of the simulations
described above with similar signal-to-noise levels. We therefore
conclude that results derived from the simulations are valid for
real data.

\medskip

\noindent{\bf 4. Conclusions}

\medskip

The major conclusion of this paper is that simple, two-layer
backpropagation neural networks can be trained to perform useful
binary classification on realistic samples of survey-quality spectral
data. Training sets should contain $\sim 100$ spectra, with noise
characteristics comparable to the data to be classified. Neural network
classification of BALQSOs and Seyfert galaxies showed success rates
of $\sim 90$\% for signal-to-noise ratios $\sim 10$, and still does
significantly better than randomly for signal-to-noise ratios as low as
$\sim 3$.

It is not necessary to know redshifts for the spectra if the classification
is done in the Fourier domain. If, however, redshift information is
available, it improves the performance of the classification when 
signal-to-noise ratios are low.

\newpage

\noindent{\bf References}

\medskip

\medskip
\noindent
Boroson, T. A., \& Green, R. F. 1992, ApJS, 80, 109

\medskip
\noindent
Francis, P. J., Hewett, P. C., Foltz, C. B., \& Chaffee, F. H. 1992, ApJ 398,
476

\medskip 
\noindent 
Francis, P. J., Hewett, P. C., Foltz, C. B., Chaffee, F. H., 
Weymann, R. J. \& Morris, S. L., 1991, ApJ, 373, 465 

\medskip
\noindent
Francis, P. J., Hooper, E. J., \& Impey, C. D. 1993, AJ 106, 417.

\medskip 
\noindent 
Gorman, R. \& Sejnowski, T. J., 1988, Neural Networks,  1, 75 

\medskip
\noindent
Hertz, J., Krogh, A., \& Palmer, R. G. 1991, in Introduction to the Theory
of Neural Computation, (Addison-Wesley)

\medskip
\noindent
Hewett, P. C., Foltz, C. B., \& Chaffee, F. H.  1995, ApJ, 445, 62

\medskip
\noindent
Hinton, G. E., 1992, Scientific American, September, 267, 3, 145 

\medskip 
\noindent 
Lahav, O., Naim, A., Buta, R. J., Corwin, H. G., de Vaucouleurs, G., 
Dressler, A., Huchra, J. P., van den Bergh, S., Raychaudhury, L., 
Sodre Jr., L., \& Storrie-Lombardi, M. C., 1994, Science, 267, 859

\medskip
\noindent
Lahav, O., Naim, A., Sodre Jr., L., \& Storrie-Lombardie, M. C., 1995 
Submitted to MNRAS

\medskip
\noindent
Maren, A. J., Harston, C. T., \& Pap, R., 1990, Handbook of 
Neural Computing (Academic Press Inc.) 

\medskip
\noindent
Osterbrock, D. E. 1989, Astrophysics of Gaseous Nebulae and Active
Galactic Nuclei (University Science Books, Mill Valley)

\medskip
\noindent
Press, W.H., Teukolsky, S.A., Vetterling, W.T. \& Flannery, B.P., 1992,
Numerical Recipes in C, (Cambridge University Press).

\medskip
\noindent
Rao, V.B. \& Rao, N.V., 1993, C++ Neural Networks and Fuzzy Logic 
(MIS:Press, New York)

\medskip
\noindent
Rawson, D. M., 1994 Neural Nets and Quasi-Stellar Objects (Honours Thesis, 
Melbourne University)

\medskip
\noindent
Rayburn, D. B., Januskiewicz, A. J., Ripple, G. R., Truwit, J., 
Summey, H. F., Young, T. R., Klimasauskas, C. C., Lee J. M., 
\& Snapper, J. R., 1991, Neural Networks, 4, 525

\medskip
\noindent
Storrie-Lombardi, M. C., Lahav, O., Sodre Jr, L., \& 
Storrie-Lombardi, L. J., 1992, MNRAS, 259, 8 

\medskip
\noindent
Taylor, K., 1995, in Wide Field Spectroscopy and the Distant Universe, 
The 35th Herstmonceux Conference, eds. Maddox, S.J. and Aragon-Salamanca, A. 
(World Scientific Publishing, Singapore) p. 15.

\medskip
\noindent
van Camp, D., 1992, Scientific American, September, 267, 3, 170 

\medskip 
\noindent 
von Hippel, T., Storrie-Lombardie, L. J., Storrie-Lombardi, M. C., \& 
Irwin, M. J., 1994, MNRAS, 269, 97 
 
\medskip 
\noindent 
Weymann, R. J., Morris, S. L., Foltz, C. B. \& Hewett, P. C., 1991, ApJ 373, 23

\newpage

\noindent
{\bf Figure Captions:}

\medskip

\noindent
{\bf Figure 1}---The success rate of the neural network classification,
as a function of noise, for the simulated Seyfert spectra. 50\% is the 
level expected by chance. The solid line is for the case when the redshift 
is known ($z=0.0$),
and the dashed line is for the case when the redshift is unknown ($0.0 < z < 
0.1$), and the Fourier technique is applied.

\medskip
\noindent
{\bf Figure 2}---The success rate of the neural network classification,
as a function of the size of the training set. Line styles
as in Fig~1; results are shown for two different noise levels.
 
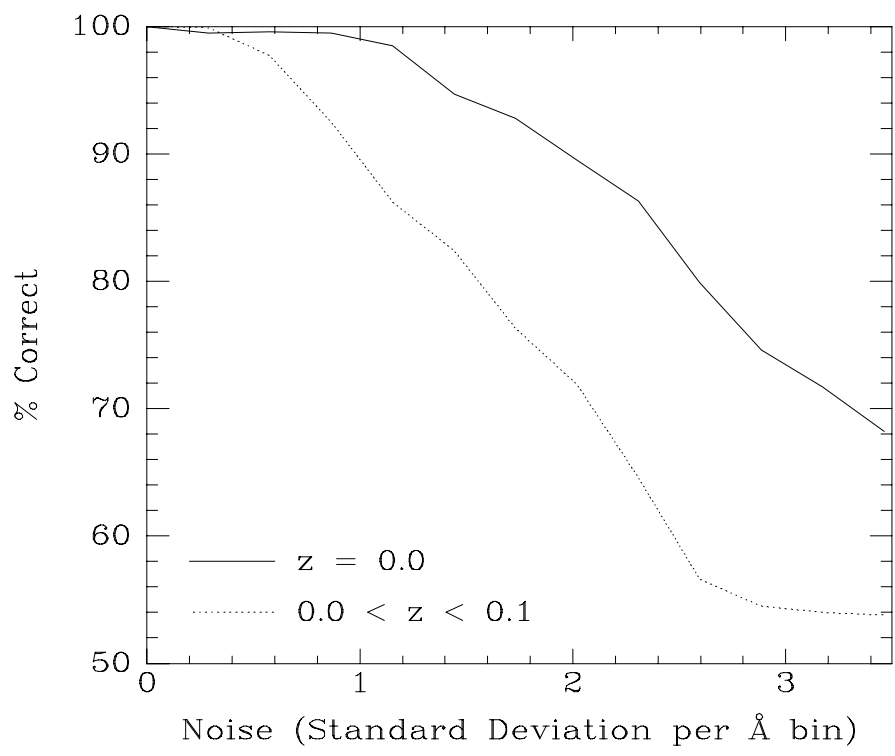
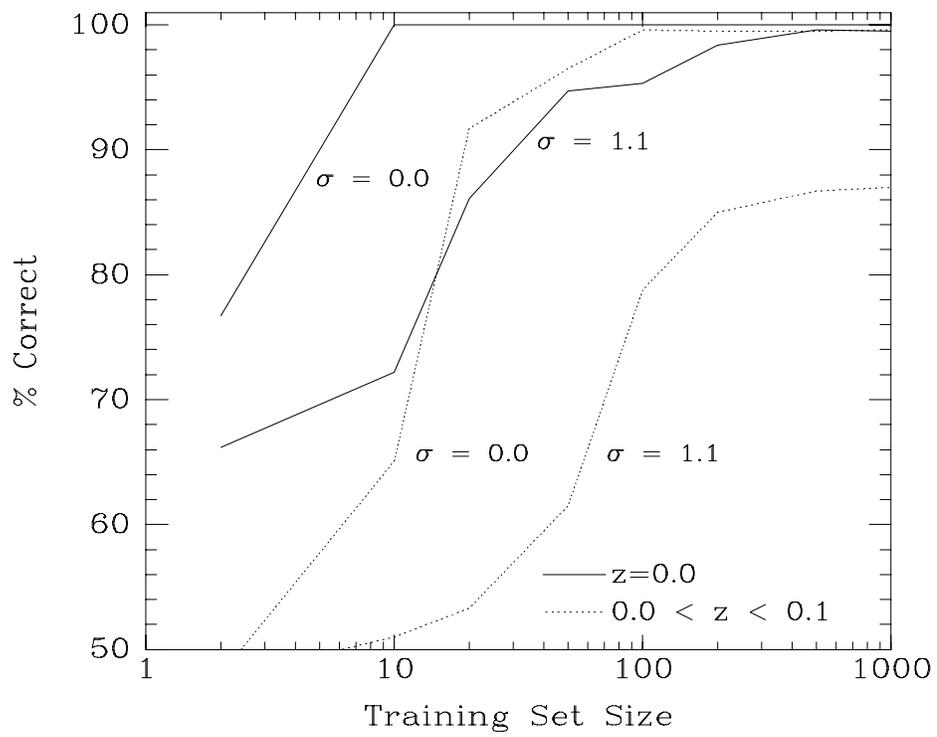
\end{document}